\documentclass[aps,prl,twocolumn,superscriptaddress,amsmath]{revtex4}
\usepackage{multirow}
\usepackage{epstopdf}
\usepackage{mathrsfs}
\usepackage{txfonts}
\usepackage{amssymb}
\usepackage{graphicx,subfigure,float}
\usepackage{dcolumn}
\usepackage{bbm}
\usepackage{bm}
\usepackage{color}
\usepackage[colorlinks, linkcolor=blue, citecolor=blue, urlcolor=blue]{hyperref}
\usepackage{lipsum}
\usepackage{gensymb}
\begin{document}
\title{Transition from antiferromagnets to altermagnets: Symmetry-Breaking Theory}
\author{P. Zhou}
\email{zhoupan71234@xtu.edu.cn}
\affiliation{Hunan Provincial Key laboratory of Thin Film Materials and Devices, School of Materials Science and Engineering, Xiangtan University, Xiangtan 411105, China}
\author{X. N. Peng}
\affiliation{Hunan Provincial Key laboratory of Thin Film Materials and Devices, School of Materials Science and Engineering, Xiangtan University, Xiangtan 411105, China}
\author{Y. Z. Hu}
\affiliation{Hunan Provincial Key laboratory of Thin Film Materials and Devices, School of Materials Science and Engineering, Xiangtan University, Xiangtan 411105, China}
\author{B. R. Pan}
\affiliation{Hunan Provincial Key laboratory of Thin Film Materials and Devices, School of Materials Science and Engineering, Xiangtan University, Xiangtan 411105, China}
\author{S. M. Liu}
\affiliation{Hunan Provincial Key laboratory of Thin Film Materials and Devices, School of Materials Science and Engineering, Xiangtan University, Xiangtan 411105, China}
\author{P. B. Lyu}
\affiliation{Hunan Provincial Key laboratory of Thin Film Materials and Devices, School of Materials Science and Engineering, Xiangtan University, Xiangtan 411105, China}
\author{L. Z. Sun}
\email{lzsun@xtu.edu.cn}
\affiliation{Hunan Provincial Key laboratory of Thin Film Materials and Devices, School of Materials Science and Engineering, Xiangtan University, Xiangtan 411105, China}
\date{\today}
\begin{abstract}
Considering the similarity of the real-space configurations for the opposite spin sublattices in both antiferromagnets (AFM) and altermagnets (AM), the relationship between them should be profound. In this work, we demonstrate that AFM and AM can be connected with spin groups and their subgroups. Consequently, the breaking of the combined inversion or translation operation with time-reversal symmetry ($PT$ or $\textbf{\emph{t}}T$) in AFM will induce transition from AFM to AM. We systematically list all collinear spin point groups and space groups that can realize the transition for the three types of AFMs: $PT$-type, $\textbf{\emph{t}}T$-type and $PT$-$\textbf{\emph{t}}T$-type. Moreover, we propose that Floquet engineering using circularly polarized light and surface cutting engineering are effective approaches to break $PT$ and $\textbf{\emph{t}}T$ symmetries of AFM, respectively, achieving the transition. Interestingly, the features and magnitude of altermagnetic spin splitting can be tuned by adjusting various parameters of Floquet engineering. Our work not only establishes a theoretical framework for the transition from AFM to AM, but also provides practical approaches utilizing the achievements in AFM for a hundred years to obtain AM, significantly expanding the scope of altermagnetic materials for both theoretical studies and future practical applications.\\
\end{abstract}
\maketitle
\indent \emph{Introduction}---Recently, altermagnets (AMs) has drawn significant interest from the scientific community following its theoretical proposal and experimental confirmation\cite{alter1,alter2,alter3,alter4,alter5,alter6,alter7,RuO21,RuO22,RuO23,spin1,spin2, spin3,spin4,spin5,spin6,spin7,superconductivity4,MnTe}. In the new magnetic phase, its local magnetic moments in real space exhibit antiferromagnetic coupling being akin to that of antiferromagnets (AFMs), whereas its bands demonstrate alternate spin-polarization in reciprocal space. Unlike conventional AFMs whose opposite spin sublattices are connected through translation or inversion symmetries, in AMs, the opposite spin sublattices are linked by crystal-rotation symmetries, potentially augmented by additional translation or inversion operations\cite{alter1}. Although the different symmetry requirements result in distinct spin groups to describe antiferromagnetic and altermagnetic systems, considering the similarity of real space configuration of the opposite spin sublattices in AFM and AM, the two types of spin groups may be connected by group-subgroup relationship. Namely, the spin group of AM is one of the subgroups of AFM. If the connection is to be built, we can obtain the AM from AFM through symmetry breaking. Such transition will not only significantly expand the scope of altermagnetic materials based on the achievements in AFM for a hundred years, but also provide an effective modulation approach for antiferromagnetic spintronics\cite{antiferromagnetic1, antiferromagnetic2}. However, there is still a lack of systematic exploration of the symmetry connection and symmetry breaking methods between the two magnetic phases.\\
\indent Until now, several primary approaches for converting AFMs into AMs have been proposed that can categorized as: van der Waals (vdW) stacking engineering of two-dimensional (2D) materials\cite{twist_1,twist_2,twist_3}, Janus structures\cite{Janus_1, Janus_2}, and external electric fields\cite{Electric_1,Electric_2,Electric_3,Electric_4}. The vdW stacking engineering involves weak magnetic exchange coupling and subsequently low N\'{e}el temperature. Janus structures will introduce extra disorder altered intrinsic structure of materials. Although electric field modulation offers a non-invasive method to break $PT$ symmetry, making it particularly promising for spintronic applications\cite{Electric_1,2D_1,2D_2}, it is limited with constraints of the directionality and symmetry breaking types. Especially, all the above methods are primarily applicable to 2D materials. However, experimentally realized 2D antiferromagnetic materials remain relatively scarce, whereas three-dimensional (3D) antiferromagnetic materials are quite abundant, such as tetragonal CuMnAs\cite{CuMnAs}, cubic NiO\cite{NiO}, and $G$-type Cr$_2$O$_3$\cite{Cr2O3}, among others. Furthermore, general understanding and approach for the transition from AFM to AM are still lack. Therefore, developing theoretical framework of the transition from AFM to AM and effective methods to induce the magnetic transitions in both 2D and 3D collinear antiferromagnets are essential for the research advancement in this field.\\
\indent In this work, we systematically analyze the symmetry relationships between collinear AFMs and AMs through group-subgroup relationships and identify all possible magnetic transition pathways driven by symmetry breaking. For $PT$-symmetry related AFMs ($PT$-type AFM), we identify 18 point group-subgroup pairs corresponding to 82 transition pathways, encompassing 252 spin space groups describing antiferromagnetic systems to 422 spin space groups describing altermagnetic systems. There are 155 and 163 spin space group for $\bm{t}T$-symmetry connected AFMs ($\bm{t}T$-type AFM) and AFMs possessing both $PT$ and $\bm{t}T$ symmetries (classified as $PT$-$\bm{t}T$-type AFM), respectively, can realize such transition. We propose two practical methods: Floquet engineering and surface engineering to induce such transitions via $PT$- and $\bm{t}T$-symmetry breaking. The magnetic transitions by breaking $PT$ and/or $\bm{t}T$ symmetry are thoroughly discussed with tight-binding models and confirmed with first-principles calculations with typical real AFMs Cr$_2$O$_3$, CaFeO$_2$, KMnP, and NiO.\\
\begin{figure}
	\centering
	\includegraphics[trim={0.0in 0.0in 0.0in 0.0in},clip,width=0.9\linewidth]{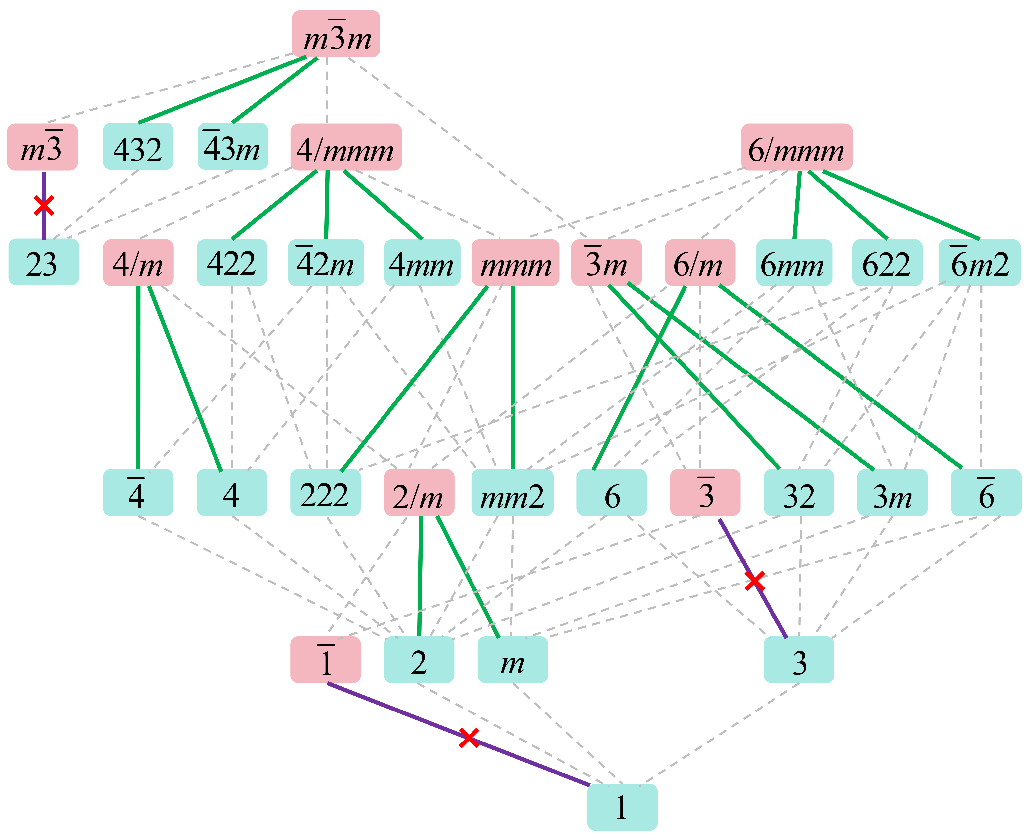}
	\caption{The evolutionary relationship between point groups and their subgroups. The point groups with inversion symmetry are denoted by pink boxes, while those without inversion symmetry are represented by arctic blue boxes. Transition pathways from groups with inversion symmetry to their subgroups without it are indicated by both green and purple lines. However, only the green lines may describe the transition from antiferromagnetic to altermagnetic phases.}\label{fig1}
\end{figure}
\indent \emph{Symmetry analysis}---Here, we focus on the transition between collinear AFM and AM due to the collinear nature of altermagnetism\cite{alter1,alter2,alter_mat}. In conventional collinear antiferromagnetic materials, opposite spin sublattices are connected by either space inversion or/and translation with time-reversal symmetry, namely AFMs can be categorized by $PT$-type AFMs, $\bm{t}T$-type AFMs, and $PT$-$\bm{t}T$-type AFMs. However, in altermagnetic materials, the opposite spin sublattices are connected by symmetry operations other than space inversion or pure translation (we name the operation as $AT$). Therefore, the key to the transition from AFMs to AMs lies in breaking space inversion or/and translation symmetry while preserving other symmetries to connect opposite spin sublattices. It is important to noted that, in collinear AFM and AM, spin orientations are limited to two antiparallel directions and remain unchanged across magnetic transitions. Thus, time-reversal and spin-rotation symmetries are preserved, while certain combined symmetries with crystalline operations may be broken. \\
\indent We firstly study the symmetry evolution of the $PT$-type AFM with the point group symmetry of the crystal. When spin freedom is disregarded, crystal symmetries are described by 32 point groups. The group-subgroup relationship among these 32 point groups is shown in Fig. \ref{fig1}, where pink and arctic blue boxes represent point groups with and without space inversion symmetry, respectively. As shown in Fig. \ref{fig1}, we have identified 18 potential pathways (indicated by green lines) that break space inversion symmetry, through which the transition from AFMs to AMs can be realized.\\
\indent The point groups alone are insufficient to fully describe the symmetries of the two types of collinear magnetic materials. It is necessary to consider the spin point groups and spin space groups\cite{sgroup1, sgroup2, sgroup3}. For the two types of magnetic states, we must use spin group $\bm{r}_s \times \mathcal{R}^{III}_{s}$, where $\bm{r}_s$ is spin only group and $\mathcal{R}^{III}_{s}$ is the third type nontrivial spin groups\cite{alter1,alter2}:
\begin{equation}\label{equ1}
\mathcal{R}^{III}_{s} = [E \Vert \bm{\mathcal{N}}] \oplus [C_2 \Vert \bm{\mathcal{G} - \mathcal{N}}],
\end{equation}
where $\mathcal{N}$ is a halving subgroup of a crystallographic point group $\mathcal{G}$, and it exclusively encompasses real-space transformations that interchange atoms within the same-spin sublattices. Whereas the coset $\bm{\mathcal{G} - \mathcal{N}}$ includes only real-space transformations that interchange atoms between opposite-spin sublattices, which can be written as $\mathcal{AN}$, where $\mathcal{A}$ denotes any operations among it. According to the labels of spin-group symmetries\cite{sgroup1, sgroup2, sgroup3, alter1}, the operations in $\mathcal{N}$ and $\mathcal{A}\mathcal{N}$, respectively, can be written as $[E \Vert A]$ and $[C_2 \Vert B]$, where the $E$ and $C_2$ in front of the double vertical line represent spin-space identity and inversion, respectively. $A$ and $B$ can be any real-space point group operations. In the labels of $[C_2 \Vert B]$, $[C_{2} \Vert P]$ and $[C_{2} \Vert \bm{t}]$, respectively, represent the conventional $PT$ or $\bm{t}T$ operation. It is worth emphasizing that, in the absence of spin-orbit coupling (SOC), type-III magnetic point groups\cite{book1} and collinear nontrivial spin groups are built in a similar manner, and both have been employed to describe antiferromagnetic or altermagnetic materials. The process of constructing of them involves selecting a crystalline point group, then identifying a subgroup of index 2. The operations of this subgroup are used to link the magnetic sublattices with the same spin direction, while the remaining half of the operations connect sublattices with opposite spins. By examining all 32 crystalline point groups, we find that 58 type-III magnetic point groups or 58 nontrivial spin groups can be derived.\\
\indent We firstly discuss the case of breaking $[C_2 \Vert P]$ symmetry in nontrivial spin groups\cite{sgroup1, sgroup2, sgroup3, alter1}. For a spin point group that used to describe antiferromagnetic materials, the space inversion operation $P$ must be included in $\mathcal{A}\mathcal{N}$. However, for altermagnetic materials, $P$ cannot be included in $\mathcal{A}\mathcal{N}$; instead, at least one rotation or rotation-inversion combined operation must be presented in $\mathcal{A}\mathcal{N}$. Therefore, if $\mathcal{A}\mathcal{N}$ for a spin point group contains both $P$ and other rotation or rotation-inversion operations, breaking the inversion operation $P$ while preserving the rotation or rotation-inversion operations should guarantee a transition from AFMs to AMs. If we disregard the transformation of spin freedom, it is sufficient to analyze the relationship between different point groups. As shown in Fig. \ref{fig1}, there are 21 such possible pathways to break the inversion operation in the point group. However, three pathways denoted by purple lines will not preserve any operations in the coset. Therefore, there are only 18 potential pathways denoted by the green lines in Fig. \ref{fig1} for the transition from AFMs to AMs. The 18 possible pathways achieving the space inversion symmetry breaking correspond to 9 original crystalline point groups, which correspond to 19 collinear spin point groups possessing inversion symmetry $P$ in the coset $\mathcal{A}\mathcal{N}$, as shown in the Fig. S1-S10. By analyzing all their subgroups that breaking the inversion symmetry, we identified 82 potential pathways for the magnetic transitions from antiferromagnetic to altermagnetic phase, as consolidated in Tab. S2 of supplemental material\cite{sup}. It is worth noting that if inversion operation is included in $\mathcal{N}$ in an altermagnetic material, it cannot be derived from corresponding antiferromagnetic one. Like the $PT$-symmetric counterpart, $\bm{t}T$-type and $PT$-$\bm{t}T$-type AFM can undergo a magnetic transition to an AM through the breaking of $\bm{t}T$ or/and $PT$ symmetries. Our in-depth analysis of the group-subgroup relationship shows that the $\bm{t}T$-type AFM transition can be realized by 155 spin space groups, while the $PT$-$\bm{t}T$-type AFM transition can be realized by 163 spin space groups. Further details are available in the Sec. I of supplemental material\cite{sup}.\\
\begin{figure}
	\centering
	\includegraphics[trim={0.0in 0.0in 0.0in 0.0in},clip,width=\linewidth]{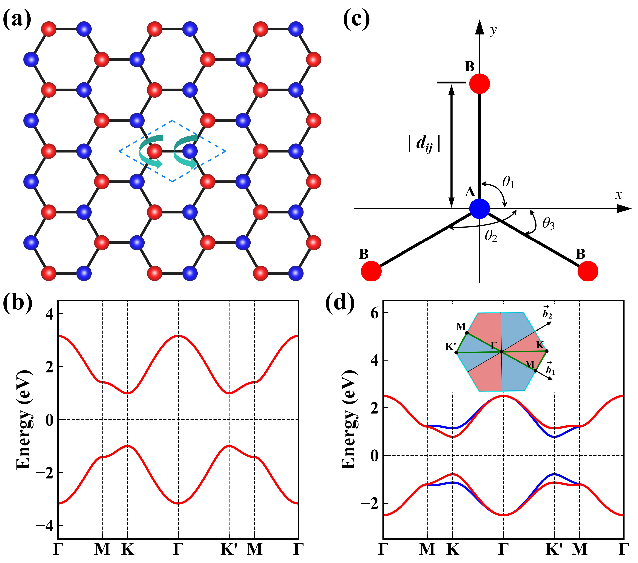}
	\caption{AFMs to AMs by Floquet engineering in $PT$ antiferromagnetic honeycomb lattice. (a) Antiferromagnetic honeycomb lattice. Blue and red represent the spin-up and spin-down sublattices, respectively. (b) The spin-degenerate energy bands in antiferromagnetic honeycomb lattice. (c) Floquet hopping of the honeycomb lattice. (d) The energy band structure with left circularly polarized light intensity 1.0/\AA. The inset illustration shows 2D Brillouin zone (BZ) with altermagnetic spin splitting. Blue and red represent spin-up and spin-down, respectively.}\label{fig2}
\end{figure}
\indent \emph{Floquet engineering}---We then investigate general and feasible methods breaking $PT$ or/and $\bm{t}T$ symmetries to induce the magnetic transition as mentioned above. Regarding the $PT$ symmetry breaking, applying a vertical electric field is an effective method to induce the magnetic transition, as discussed in detail for antiferromagnetic MoTe in Sec. II of the supplemental material\cite{sup}. However, this approach is limited to 2D materials. Here we propose Floquet engineering is a general and feasible approach suitable for both 2D and 3D antiferromagnetic materials to realize such transition. Floquet engineering is a sophisticated technique in quantum physics and condensed matter systems, which utilizes time-periodic perturbations to dynamically control and manipulate the properties of materials. Recent studies have highlighted its ability to control the electronic, magnetic, and topological properties of materials\cite{fapp_1,fapp_2,fapp_3}. In this work, we find that applying circularly polarized light is a general approach to break the $PT$ symmetry in AFMs, facilitating its magnetic transition to AM.\\
\indent To demonstrate the symmetry-breaking mechanism of Floquet engineering, we use the antiferromagnetic honeycomb structure as an example, as shown in Fig. \ref{fig2}(a). In the absence of an external field, its Hamiltonian can be expressed as
\begin{equation}\label{equ2}
\mathcal{H} = t \sum_{\langle i, j\rangle,\sigma} c_{i,\sigma}^{\dagger} c_{j,\sigma} - J_{sd} \sum_{i,\sigma,\sigma^{\prime}} S_{i} \cdot c_{i,\sigma}^{\dagger} \sigma_{\sigma \sigma^{\prime}} c_{i,\sigma^{\prime}} + h.c.,
\end{equation}
the term $c_{i,\sigma}^{(\dagger)}$ represents the annihilation (creation) operator for an electron at site $i$ with spin $\sigma$. The nearest-neighbor hopping strength is denoted by $t$, and the coupling constant $J_{sd}$ characterizes the on-site exchange interaction between the spins of the itinerant electrons and the localized spins at the magnetic sites $S_{i}$. The antiferromagnetic honeycomb lattice is described by the spin point group $^{2}6/^{1}m^{2}m^{1}m$. Due to the protection of $PT$ symmetry, the spin-up and spin-down bands remain degenerate throughout the whole BZ, as shown in Fig. \ref{fig2}(b). When left circularly polarized light, described by $\textbf{A}(t) = A[ \cos(\omega t), - \sin(\omega t), 0]$ (with photon energy \( \hbar \omega = 10 \, \text{eV} \)) is applied to the system, the hopping interactions are modified by an additional phase factor $\exp\left(i \frac{e}{\hbar} \textbf{A}(t) \cdot \textbf{d}_{ij}\right)$, where \( \textbf{d}_{ij} \) is the relative position vector between orbitals \( i \) and \( j \). According to Floquet-Bloch theory under the high-frequency approximation \cite{Floque_1,Floque_2,Floque_3,Floque_4}, this time-dependent modulation can be effectively replaced by a time-independent term of the form $\exp(-i q \theta) \cdot i^{q} J_q \left(\frac{e}{\hbar} A\right)$, where \( \theta \) is the angle between \( \textbf{d}_{ij} \) and the \( x \)-axis [see Fig. 2(c)], and \( J_q \) denotes the Bessel function of the first kind of order \( q \) (a more detailed derivation can be found in the supplemental material \cite{sup}). As a result, the effective Floquet hopping modulated by the light field breaks the \( PT \) symmetry of the system. This process can also be viewed as applying a pseudomagnetic vector of the same direction to the opposite spin sublattice, as shown in Fig. \ref{fig2}(a). Detailed symmetry analysis reveals that the system breaks the $PT$ and \( [C_2{\parallel}C_{6z}] \) symmetries, and maintains the \( [E{\parallel}C_{3z}] \) and \( [E{\parallel}M_z] \) symmetries, i.e., the spin point group changes to \( ^{1}\bar{6}^{2}m^{2}2 \). As shown in Fig. 2(d), the energy bands exhibit the typical characteristics of altermagnetic spin splitting, with opposite spin sublattices connected by three in-plane \( C_2 \) or three vertical mirror symmetries. It is important to note that if the left circularly polarized light is replaced with right circularly polarized light, the spin splitting can be reversed, as illustrated in Fig. S20(b) of supplemental material. Additionally, the magnitude of spin splitting can be adjusted by altering the light intensity. For example, the spin splitting at the K point can increase from 0.356 eV to 0.454 eV when the value of $A$ changes from 1.0$/\AA^{-1}$ to 1.2$/\AA^{-1}$, as shown in Fig. S20(a) and S20(d). These findings suggest that through Floquet engineering, we can achieve spin-tunable altermagnetic materials. Moreover, if we remove the light field, the system would restores the equilibrium antiferromagnetic state. This dynamic control provides an ideal platform for the ultrafast and reversible generation of spin currents. Therefore, our general theory and practical methods not only expand the class of altermagnetic materials to include traditional high N\'eel temperature AFMs, but also pave the way for efficient and ultrafast control of spintronic functionalities.\\
\begin{figure}[t]
	\centering
	\includegraphics[trim={0.0in 0.0in 0.0in 0.0in},clip,width=\linewidth]{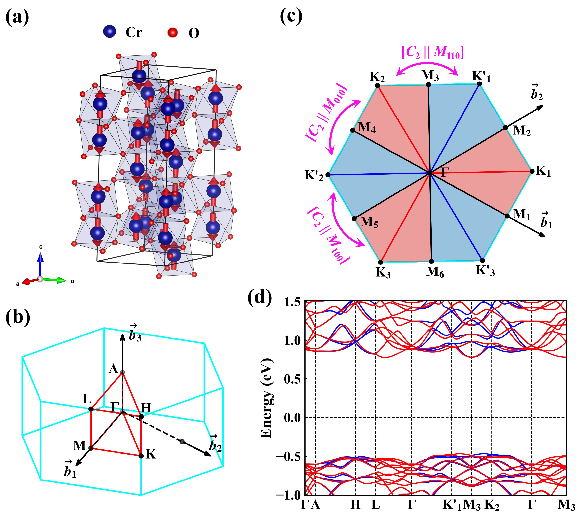}
	\caption{AFMs to AMs by Floquet engineering in $G$-type antiferromagnet Cr$_2$O$_3$. (a) and (b) Crystal structure and the 3D BZ of the $G$-type AFM Cr$_2$O$_3$, respectively. (c) The BZ plane at $k_z = 0$ exhibits altermagnetic spin splitting. (d) Energy bands under left circularly polarized light ($\hbar \omega$ = 15 eV) with an intensity of 0.35/{\AA}, where blue and red denote spin-up and spin-down states, respectively.}\label{fig3}
\end{figure}
\indent To confirm the symmetry breaking of Floquet engineering on $PT$-type AFM, we take $G$-type antiferromagnetic Cr$_2$O$_3$ as a typical example. Cr$_2$O$_3$ is notable for its exceptionally high antiferromagnetic ordering temperature of approximate 300 K \cite{Cr2O3_2}, and it has been drawn considerable attention for its intriguing magnetoelectric properties \cite{Cr2O3_3}. As depicted in Fig. \ref{fig3}(a), the adjacent magnetic Cr atoms in each vertical row show opposite spins, exhibiting the characteristics of $G$-type antiferromagnetic order. The symmetries of the material can be described by the spin point group $^{2}\bar{3}^{2}m$. The energy bands of Cr$_2$O$_3$ remain spin-degenerate across the entire BZ, as shown in Fig. S22(b), due to the presence of $PT$ symmetry. When left circularly polarized light is applied along the $z$ direction, the $PT$ symmetry of the system is broken, leading to altermagnetic spin splitting, as illustrated in Fig. \ref{fig3}(c) and \ref{fig3}(d). Furthermore, the light field induces pseudomagnetic vectors, disrupting the in-plane $C_2$ symmetry. The spin point group of Cr$_2$O$_3$ now change to be $^{1}3^{2}m$. Consequently, the opposite spin sublattices of Cr$_2$O$_3$ are now connected by three vertical mirror symmetries: $[C_2 \Vert M_{100}]$, $[C_2 \Vert M_{010}]$, and $[C_2 \Vert M_{110}]$, as shown in Fig. \ref{fig3}(c). The three vertical mirror symmetries will protect the formation of three vertical spin-degenerate nodal planes within the 3D BZ. It is worth to mention that Cr$_2$O$_3$ is a typical 3D materials. For 2D materials, Floquet engineering is also suitable for inducing the transition from AFM to AM through $PT$ symmetry breaking. A typical example of 2D MnSr can be found in Sec. III of the supplemental material\cite{sup}.\\
\indent \emph{Surface engineering}---In addition to $PT$-type AFM, $\bm{t}T$-type AFM is also a large class of traditional AFMs ($\bm{t}$ is a fractional translation along the lattice basis vector direction, and here we assume without $PT$ symmetry in the $\bm{t}T$-type AFM connecting the opposite spin sublattices). If SOC is ignored, their symmetries must be described by spin space groups. For this type of AFM, we propose that surface engineering can convert them into AM. After slicing a surface to intercept the fractional translation of the $\bm{t}T$-type AFM, an interface is formed with the vacuum along the Miller plane ($hkl$). We define $\hat{\textbf{n}}$ as the unit vector in the normal direction of the surface and it is a key parameter to characterize the surface symmetry. As discussed in the S. F. Weber et al's work\cite{surfmag2}, the surface symmetry group is a subgroup of the group of the bulk. Here we label it as $\textbf{G}_{\hat{\textbf{n}}} = [\textbf{R}_{i} \Vert (\textbf{R}^{\hat{\textbf{n}}}_{j} \vert \bm{t}^{\perp \hat{\textbf{n}}})]$, as shown in Fig. S23. Because we concentrate on the surface that host altermagnetism, we only study the surface hosting equal opposite spin sublattices that the spin compensating with each other. Such surfaces are different from the surface with equilibrium magnetization in previous reports\cite{surfmag1,surfmag2}. Provided no $\bm{t}T$ operations connect the opposite spin sublattices in the surface, while other compound operations linking the opposite spin sublattices, the altermagnetism will emerge. By analyzing all 289 spin space groups of $\bm{t}T$-type AFM, we find 155 among them host altermagnetic-like subgroups, namely the systems with these spin space groups can be potentially turn into AMs with $\bm{t}T$ symmetry breaking.\\
\indent To illustrate the transition driven by $\bm{t}T$ symmetry breaking in the $\bm{t}T$-type AFM class, we consider the pure $\bm{t}T$ AFM CaFeO$_2$. This compound crystallizes in a layered structure with a high-spin, distorted square-planar configuration, and exhibits $G$-type AFM ordering with a N\'eel temperature of 420 K\cite{CaFeO2}. The magnetic moments of the Fe atoms exhibit AFM coupling, forming an in-plane magnetic order within the (001) plane. Additionally, antiparallel alignment of Fe spins along the $c$-axis leads to a doubling of the lattice constant $c$ in the magnetic unit cell, as depicted in Fig. S24(a). In this compound, the two spin sublattices are related by the combined symmetry operation $\bm{t}T$, where $\bm{t} = (0, 0, \frac{1}{2})$. The corresponding spin space group is given by $G_{ss} = {P}{c}^{1}{\bar{4}}^{1}{2_{1}}^{1}c^{{\infty}m}1$. The electronic band structure of CaFeO$_2$ displays semiconducting behavior with a band gap of 1.80 eV. Due to the presence of $\bm{t}T$ symmetry, the band structure is entirely spin-degenerate throughout the BZ, as shown in Fig. S24(b). To break the $\bm{t}T$ symmetry, we construct a slab geometry with its surface normal oriented along the $\bm{t}$ direction. This breaks the $\bm{t}T$ symmetry, resulting in a change in the spin space group to ${P}^{1}{\bar{4}}^{-1}{2_{1}}^{-1}m^{{\infty}m}1$. In this group, the symmetries $[C_2 \Vert M_{110}]$ and $[C_2 \Vert M_{1\bar{1}0}]$ are preserved. As a consequence, spin degeneracy is lifted except along the $\tilde{\Gamma}-\tilde{M}$ high-symmetry line of 2D surface BZ, as shown in Fig. S24(d). This leads to a characteristic $d$-wave altermagnetic spin distribution, illustrated in Fig. S24(c).\\
\begin{figure}[t]
	\centering
	\includegraphics[trim={0.0in 0.0in 0.0in 0.0in},clip,width=\linewidth]{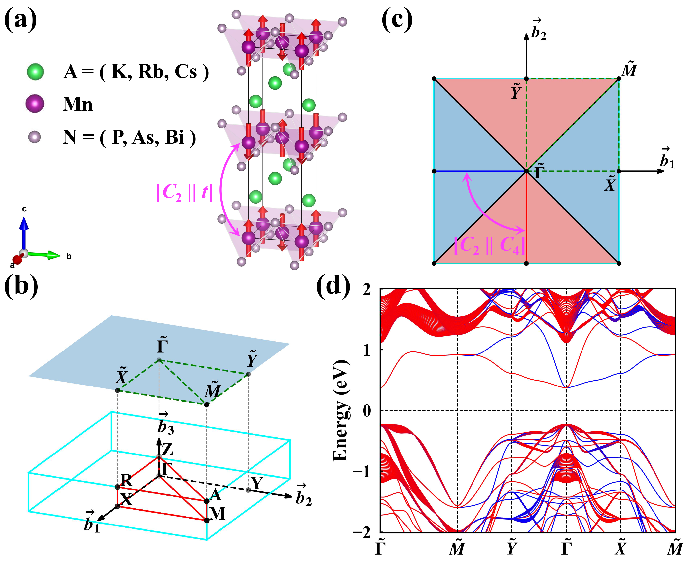}
	\caption{Surface engineering-induced transition from AFMs to AMs in $\bm{t}T$ antiferromagnetic materials AMnN (A = K, Rb, Cs; N = P, As, Bi). (a) Crystal structure and (b) 3D Brillouin zone (BZ) of AMnN. (c) Surface 2D BZ and altermagnetic spin splitting. (d) Slab band structure of KMnP with 15 layers along the $z$-direction, where blue and red represent spin-up and spin-down states, respectively.}\label{fig4}
\end{figure}
\indent In addition to $PT$-type and $\bm{t}T$-type AFMs, there are AFMs whose opposite spin sublattices are simultaneously connected by both $PT$ and $\bm{t}T$ symmetries, named as $PT$-$\bm{t}T$-type AFM. To induce the transition from $PT$-$\bm{t}T$-type AFM to AM, both $\bm{t}T$ symmetry and $PT$ symmetry must be simultaneously broken. There are a total of 228 spin space groups of this type AFM, among which 163 have altermagnetic-like subgroups that can potentially turn into AMs. Here we use a class of alkali metal manganese pnictides $PT$-$\bm{t}T$-type AFM AMnN\cite{KMnP} (A = K, Rb, Cs; N = P, As, Bi) to illustrate the phase transition through surface engineering. The magnetic moments of Mn atoms in this type of compounds exhibit antiferromagnetic coupling in the $x$-$y$ plane. Moreover, Mn atoms are also antiparallel in the $c$-axis direction leading to a doubling of the lattice constant $c$ of the magnetic unit cell as shown in Fig. \ref{fig4}(a). In the materials, the opposite spin sublattices are connected by $\bm{t}T$ [$\bm{t} = (0, 0, \frac{1}{2})$] and $PT$. The spin space group of the system is $G_{ss} = P_{c}^{1}4_{2}/^{1}n^{1}m^{1}c\,^{{\infty}m}1$. We take KMnP as a specific example, whose electronic band structure exhibits the characteristics of a semiconductor with a band gap of 1.34 eV. Due to the coexistence of $PT$ and $\bm{t}T$ symmetry, its band structure is completely spin-degenerate in reciprocal space, as shown in Fig. S22(d) of supplemental material\cite{sup}. To break its $\bm{t}T$ and $PT$ symmetries, we construct a slab structure with its surface oriented perpendicular to the $\bm{t}$ direction. The spin space group of the material changes to $P^{-1}4/^{1}n^{-1}m^{1}m\,^{{\infty}m}1$. In this spin space group, both of $\bm{t}T$ and $PT$ are broken, while the [$C_2||C_4$] symmetry is preserved, leading to spin splitting, except along the $\tilde{\Gamma}-\tilde{M}$ high symmetry line in the 2D surface Brillouin zone. This results in $d$-wave altermagnetic spin distribution, as illustrated in Fig. \ref{fig4}(c). \\
\indent To illustrate another type of transition from $PT$-$\bm{t}T$-type AFM to AM, we take the AFM NiO\cite{NiO} as an example. In the paramagnetic phase, the symmetry of rocksalt NiO belongs to cubic space group $Fm\bar{3}m$. While in the antiferromagnetic phase, the local magnetic moment of Ni atoms aligns along the [11$\bar{2}$] direction\cite{NiO_2}, and the symmetry of its magnetic primitive cell is belong to spin space group $C_{c}^{1}2/^{1}c\,^{{\infty}m}1$. The crystal structure, BZ, and high-symmetry $k$-paths of monoclinic NiO can be found in Fig. S25(a) and S25(b). The $PT$ and $\bm{t}T$ symmetries force the energy bands spin-degenerate throughout the BZ, as shown in Fig. S25(c). To break the $\bm{t}T$ symmetry, we create a 2D antiferromagnetic thin film of NiO through surface engineering, as depicted in Fig. S25(a). The spin space group of 2D antiferromagnetic NiO is $C^{-1}2/^{1}m\,^{{\infty}m}1$ that still retains $PT$ symmetry resulting in spin-degenerate electronic energy bands, as shown in Fig. S25(e). When we further apply left circularly polarized light along the [010] direction, the spin space group of the 2D NiO transforms into $C^{-1}2\,^{{\infty}m}1$. In this group, the opposite spin sublattices are only connected by $[C_2 \Vert C_{2_{010}}]$, leading to altermagnetic spin splitting. As illustrated in Figs. S25(d) and S25(f), the electronic bands of 2D NiO exhibit typical altermagnetic spin splitting under the influence of the light field.\\
\indent \emph{Discussions}---In summary, we develop a general symmetry-breaking theoretical framework and a universal implementation strategy via Floquet and surface engineering for the transition from AFM to AM. The theory and approaches offer a versatile and highly controllable platform for broad implications for both fundamental research and spintronic applications. Firstly, the theoretical framework for the phase transition from AFM to AM provides practical approaches utilizing the achievements in AFM for a hundred years to obtain AM, significantly expanding the scope of altermagnetic materials for both theoretical and experimental studies. Secondly, both Floquet and surface engineering provide a broad range of tunable abilities of the obtained AM systems, including polarization, photon energy, light amplitude, and incident orientations of periodic light fields and symmetry-dependent effects via crystallographic orientation of surface engineering. Finally, the periodic light fields not only provide a general $PT$ symmetry breaking method, but also contribute a feasible control scheme for ultrafast spin dynamics through switching the circular polarization of the light. Therefore, our general theoretical framework and practical modulation methods not only expand the class of altermagnetic materials  include traditional high N\'{e}el  temperature AFMs, but also pave the way for efficient and ultrafast control of spintronic functionalities.\\
\indent In particular, our framework goes beyond prior studies that mainly stressed the distinction between the symmetry descriptions of AFMs and AMs. Instead, we articulate the underlying connection between the two: through suitable symmetry-breaking strategies, a transition from antiferromagnetism to altermagnetism can be systematically realized. To this end, we propose two universal and experimentally relevant implementation routes---Floquet engineering and surface engineering---that enable AFM-to-AM phase transitions, with the notable advantage of being applicable to 3D materials. Importantly, our symmetry classification encompasses all three fundamental types of AFMs ($PT$-type, $\bm{t}T$-type, and $PT$-$\bm{t}T$-type). This unified theoretical framework, together with practical modulation strategies, not only extends the family of altermagnetic materials by incorporating conventional high-N\'{e}el-temperature AFMs, but also provides a foundation for efficient and ultrafast control of antiferromagnetic spintronics. Crucially, the transition to the altermagnetic phase offers tunable physical responses inaccessible in the parent antiferromagnetic state, such as the anomalous Hall effect, spin current generation, and nonlinear transport, thereby pointing toward concrete experimental realizations of new physics beyond symmetry cataloging.\\
\indent To realize optic-field-induced altermagnetic Floquet-Bloch states in antiferromagnetic materials, a sufficiently strong light-matter interaction is critical to drive the formation of these states. Recent experimental evidence demonstrates that increasing the laser intensity effectively enhances the light-matter coupling strength, with typical electric field strengths ranging from 0 to 1 V/{\AA}. However, excessively high intensities may induce heating or nonlinear absorption, potentially damaging the material's intrinsic properties. To mitigate this, optical cavities can be employed to significantly amplify the effective Floquet field at extremely low fluences, as demonstrated in recent studies\cite{c_floquet}. Furthermore, the interaction between light and the material must account for the optical penetration depth. Given that recent advancements in Floquet engineering primarily focus on layered materials with finite thickness, selecting antiferromagnetic materials of this type is crucial for observing the transition from antiferromagnetic to altermagnetic states under optical driving. The material thickness should be carefully optimized-neither too thin nor too thick-with experimental results suggesting an optimal range of 5-20 nm to balance light penetration and surface effects, thereby enhancing the observability of Floquet-induced states.\\
\indent \emph{Acknowledgments}---This work is supported by the National Natural Science Foundation of China (Grant No. 12574070 and Grant No. 1250042757), the Postgraduate Scientific Research Innovation Project of Hunan Province (CX20240616), the China Postdoctoral Science Foundation(2025M773383 and GZC20252231), the China Postdoctoral Science Foundation-Hunan Joint Support Program(2025T002HN), the Excellent Youth Funding of Hunan Provincial Education Department (22B0175), Natural Science Foundation of Hunan Province (No. 2023JJ30572).\\
\indent Pan Zhou, Xiaoning Peng and Yuzhong Hu contributed equally to this work.
\bibliography{references}
\end{document}